# Dynamics and structure of interfacial crack front

Knut Jørgen Måløy[1], Renaud Toussaint[1], and Jean Schmittbuhl[2]
[1]Department of Physics, University of Oslo, P.O.Box 1048 Blindern, N-0316 Oslo, Norway.
[2]Laborato1re de Geologie, UMR 8538 Ecole Normale Superieure,
24 rue Lhomond, 75231 Paris Cedex 05, France

ABSTRACT
The propagation of an interfacial crack front through a weak plane of a transparent Plexiglas block has been studied experimentally. A stable crack in mode I was generated by loading the system by an imposed displacement. The local velocities of the fracture front line have been measured by using an high speed CCD camera. The distribution of the velocities exhibits a power law behavior for velocities larger than the average front velocity $<v>$ with a crossover to a slowly increasing function for velocities lower than $<v>$. The fluctuations in the velocities reflect an underlying irregular bursts activity with a power law distribution of the bursts. We further found that the size of the local bursts scales differently in the direction parallel to and perpendicular to the fracture front.

## 1 INTRODUCTION

The propagation of a crack front through a heterogeneous solid is a central question for numerous mechanical problems. The scaling properties of the morphology of brittle cracks manifest themselves through self-affine [1-2] long range correlations [1-11] with a roughness exponent which is found to be very robust for different materials and a broad range of length scales [3-9]. In order to shed light on both the dynamics, the structure and the "universality" [4] of the roughness exponent a simpler problem than the original 3D one has been devised [10-11]. This problem is simpler because the crack front is constrained geometrically to lie in the plane and is driven by the stress field transmitted through two elastic plates. So far most experiments on fracture front lines have been focused on the morphology of the fracture front and less on the detailed dynamic. Recent experimental studies of the in plane fracture problem presented here gave the estimate of the roughness exponent $\zeta=0.55\pm0.05$ [10], and was followed up by a longer study leading to the estimate $\zeta=0.63\pm0.03$ [12]. A recent study of the motion of a helium-4 meniscus along a disordered substrate - a problem related to the motion of the crack line - gave $\zeta=0.56\pm0.03$ [12]. In contrast to the experiments presented here most experiments are performed with instable fractures which exhibit fast propagation with a speed of the order of the speed of sound and a direct observation of the detailed crack front line is usually impossible. The focus of this work is to study the local dynamics of the fracture propagation. The front has in recent work [13] been found to exhibit a Family-Vicksek scaling [14] with a roughness exponent $\zeta=0.63$ and a dynamic exponent $\kappa=1.2$. The results are consistent with recent quasi static simulations [15-16] and with an elastodynamic description [17-18]. In this work we went further on in the study of the local dynamics. We show that the movement of the fracture is controlled by local bursts and that the velocity distribution exhibits a power law behavior with a characteristic speed equal the average front velocity $<v>$. The dependence of the velocity distribution on the rescaled velocity $v/<v>$ is independent on the average velocity $<v>$ for all experiments ($<v>$ is ranging from 0.36μm/s to 40μm/s). The burst distribution in space was further measured and we found that the size of the bursts scales differently in the normal and the tangential direction to the crack front line.

## 2 EXPERIMENTAL PROCEDURE

Samples were made of transparent polymethylmethacrylate (PMMA) which makes the fracture front directly observable because of the transparency of the material [10-11]. Each sample was obtained by annealing two plates of dimension 32cm × 14cm × 1cm and 34cm × 12cm × 0.4cm together at 205°C. Both plates were sand blasted on one side with 50μm steel beads before the annealing. The sand-blasting procedure introduces a randomness in the annealed surface with a cutoff in the structure on a length scale of about 50μm. We don't expect correlations in the toughness fluctuations above this length scale. The annealed surface corresponds to a weak surface which the fracture front line will propagate along. The 1cm thick plate was clamped to a stiff aluminum frame and a normal displacement was applied on boundary on the short side of the 0.4 cm thick plate. The fracture front line was observed by a microscope linked to a high speed Kodak Motion Korder Analyzer camera which records up to 500 images per second with a spatial resolution of 512×240 pixels. The experiment was performed with an average front line speed ranging from 0.36μm/s to 40μm/s. In total 8 different experiments were performed with 4367 images which gives all together 34936 fracture front lines to be analyzed.

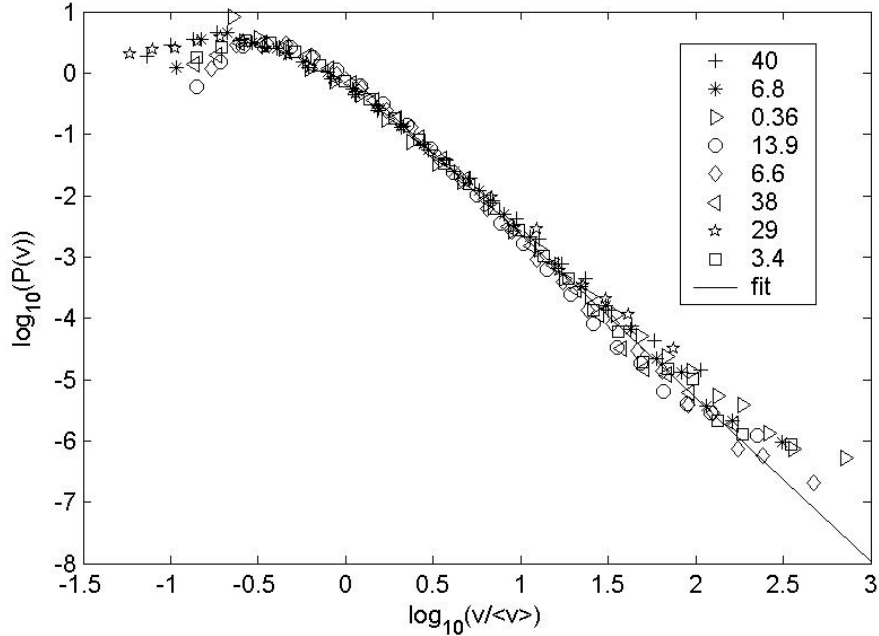

Figure 1: The velocity distribution P(v) as function of v/<v>. The solid line represents a fitting to all data for v > <v> to a linear function and has a slope -2.67. The number inserted in the figure indicates the average speed <v> in μm/s.

The fracture front lines extracted from the digital images were added to obtain a waiting time matrix **M.** The matrix **M** has the same dimension as the images and an initial value equal to zero for all its elements. The addition of the front line to the matrix was performed by adding 1 to the matrix in the positions corresponding to the front line positions. This procedure was done for all 4367 images in the experiment to obtain the final waiting time matrix **M** for each experiment. The local normal speed of the interface at the time when the front went through a particular position is found by the inverse value of the corresponding matrix element of **M** multiplied by K, where K is the ratio between the linear size of the pixel (typical 10μm) and the time between each picture (typical 0.002s). Let **V** the image matrix representing the local speeds. It is important to mention that the image recording was performed so fast that there was basically no holes in the waiting time matrix **M** with value equal to zero (apart from below the first front and above the last front, and some few artifacts due to impurities in the sample). The **V** matrix allows to associate to each pixel representing the front in each image an estimated front velocity $v=K/m$ (where m is an integer). For each possible measured velocity $v=K/m$, the probability of v is estimated as the occurrence number of this velocity over all pixels in all front images.

## 3. RESULTS

The velocity distribution P(v) is shown in Fig.1 as function of $v/<v>$ where $<v>$ is the average speed of the fracture front line for the corresponding experiments. A satisfying data collapse is obtained by scaling the local velocity v with the average velocity $<v>$. A power law behavior of the velocity distribution $P(v) \propto (v/<v>)^{-\eta}$ is apparent for velocities larger than $<v>$ with a crossover at low velocities to a slowly increasing function for velocities lower than $<v>$. The solid line, obtained from a linear fit to the experimental data for $v/<v> >1$ has a slope $-\eta=-2.67\pm0.10$.

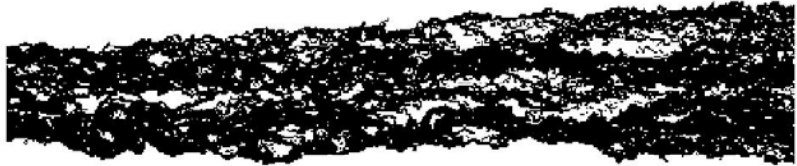

Figure 2: The image shows the distribution of bursts of size s in white for C=10 and an average velocity $<v>=29\mu m/s$.

The power law tail in the velocity distribution indicates a non trivial underlying dynamics. This dynamics can also be observed by visual inspection of the fast video recording where irregular jumps occur on all length scales. To analyze the burst activity we will consider the velocity matrix **V**. A clipped matrix was generated from **V** by setting the elements equal to one for $v>C<v>$ and zero elsewhere. Fig. 3 shows the dependence of the distribution N(s) on the connected clusters of size s in the clipped matrix for different values of C. We find a stable result for the distribution for a wide range of C values 4<C<14. A fit of the experimental data to a linear function, gives a slope -1.89±0.10. Assuming a power law behavior $N(s) \propto s^{-\gamma}$ this gives $\gamma=1.89$. As seen in Fig. 2 the typical width of the trapped islands s in the direction normal to the front are limited by the characteristic width of the fracture front line. However the width of the islands $l_1$ in the direction parallel to the front are typical larger than their width $l_2$ normal to the front. Since the fracture front is self-affine we want to check if the individual clusters exhibit the same scaling. Fig.4 shows the two distributions $n(l_1)$ and $n(l_2)$ for different clip levels C. The experimental data for both curves are consistent with power laws $n(l_1) \propto l_1^{-\alpha}$ and $n(l_2) \propto l_2^{-\beta}$. Linear fitting of the data gives $\alpha=2.3\pm0.1$ and $\beta=2.5\pm0.1$ respectively. If in addition if the self affinity of the front leads to a scaling relationship $l_2 \propto l_1^{\zeta}$, $n(l_1)dl_1=n(l_2)dl_2$ imply the following scaling law:

$$\alpha= \zeta(\beta-1)+1 \qquad (1)$$

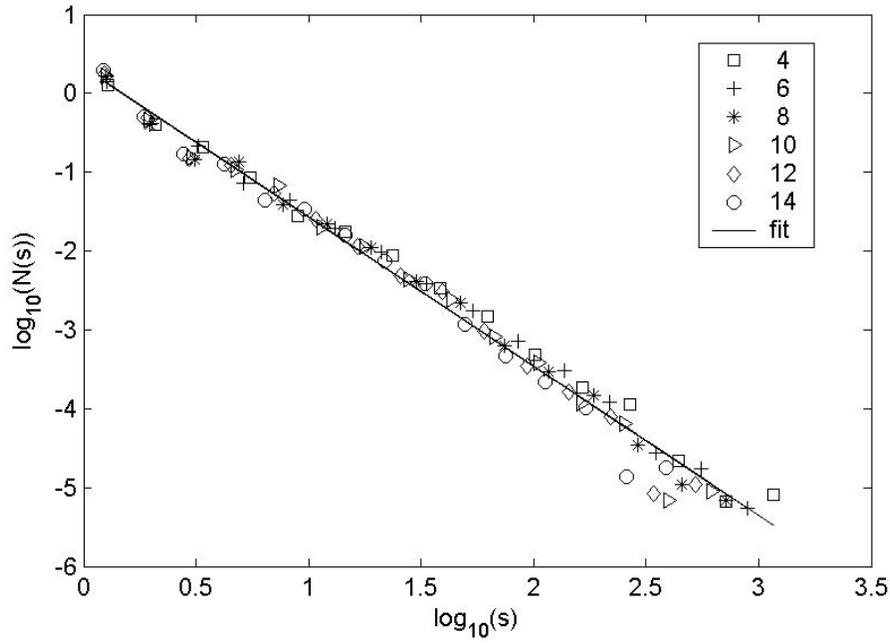

Figure 3: The distribution N(s) for different clip levels averaged over all eight experiments. The numbers inserted in the figure are the clip levels C. The solid line is a linear fitting to the data points and has a slope -1.89.

By using the fitted results of $\alpha$ and $\beta$ we find $\zeta=0.86\pm0.20$ which is somewhat higher but consistent with the roughness exponent 0.63 of the fracture front line [11]. The size of an island might as a zero order approximation be written as $s=l_1l_2$. Since $s, l_1$ and $l_2$ depend on each other we may assume $n(s)ds=n(l_1)dl_1=n(l_2)dl_2$. From this assumption follows the following scaling law between $\alpha, \beta$ and $\gamma$.

$$\gamma=(1+\beta\zeta)/(1+\zeta)=(\alpha+\zeta)/(1+\zeta)=(\alpha\beta-1)/(\alpha+\beta-2) \qquad (2)$$

By using the measured values of $\alpha$ and $\beta$ we find $\gamma=1.70\pm0.20$. This value is consistent with the value $\gamma=1.89\pm0.10$ measured directly from the burst size distribution N(s).

## 4. CONCLUSION

The fast dynamics at small scales are very different from the apparent dynamics at large scales characterized with a smooth creeping motion. We show in this work that the dynamics is controlled by an irregular burst activity with bursts of size s on all length scales but limited by the total width of the fracture front line. The bursts activity has a power law like distribution with an exponent $\gamma=1.89$. The lengths $l_1$ and $l_2$ of the bursts parallel and normal to the average front line are linked through $l_2 \propto l_1^\zeta$ with an exponent $\zeta$ consistent with the roughness exponent of the fracture front line. More experiments is however needed to be conclusive on this point due to the uncertainty in the data. The velocity distribution of the fracture front exhibit a nice data-collapse when plotted as function of $v/<v>$ and with power law behavior when $v><v>$ with an exponent $\eta=2.67$. A correct modeling of this problem should in addition to give the correct roughness exponent $\zeta$ also predict the exponent $\eta$ and $\gamma$ found in these experiments.

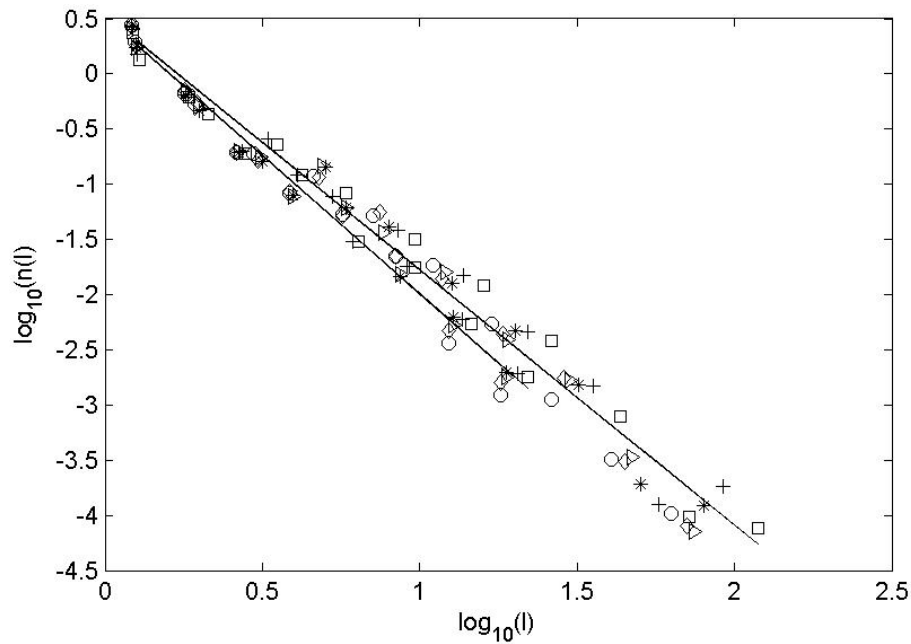

Figure 4: The upper and lower curve shows the distribution $n(l_1)$ and $n(l_2)$ respectively averaged aver all eight experiments. The solid lines are fitting to a linear curves and has slopes -2.5 and -2.3 respectively. The different symbols reflect the same clip levels as the insert in Fig. 3.

References

[1] B. B. Mandelbrot, *The fractal geometry of nature,* (W. H. Freeman New York). (1983)
[2] B. B. Mandelbrot, D. E. Passoja, and A. J. Paullay. Nature, **308,** 721, (1984)
[3] S. R. Brown and C. H. Scholz, J. Geophys. Res., **90,** 12575, (1985)
[4] E. Bouchaud, G. Lappaset, and J. Planes. Europhys Lett. **13,** 73, (1990)
[5] K. J. Måløy, A. Hansen, E. L. Henrichsen, and S. Roux. Phs. Rev. Lett., **68,** 68, (1992)
[6] J. Schmittbuhl, S. Gentier, and S. Roux, Geophys. Res. Lett., **20,** 639, (1993)
[7] B. L. Cox, and J. S. Y. Wang. Fractals, **1**, 87, (1993)
[8] W. L. Powel, T. E. Tullis, S. R. Brown, G. N. Boitnott, C. H. Scholz. Geophys. Res. Lett. **14**, 29, (1987)

[9] E. Bouchaud, J. Phys, **9,** 4319, (1997)
[10] J. Schmittbuhl, and K. J. Måløy, Phys. Rev. Lett. **78,** 3888 (1997)
[11] A. Delaplace, J. Schmittbuhl, and K. J. Måløy, Phys. Rev. **E60**, 1337 (1999)
[12] E. Rolley, C. Guthmann, R. Gombrovisch and V. Repain, Phys. Rev. Lett. **80,** 2865, 1998)
[13] K, J, Måløy, and J. Schmittbuhl, Phys. Rev. Lett. **87,** 105502, (2001)
[14] F. Family, and T. Vicsek, J. Phys. A. **18**, L75 (1985)
[15] J. Schmittbuhl, A. Hansen, and G. Batrouni, Phys. Rev. Lett. **90,** 045505 (2003)
[16] A. Hansen, and J. Schmittbuhl, Phys. Rev. Lett, Phys. Rev. Lett, **80,** 045504 (2003)
[17] S. Ramanathan and D. Fisher, Phys. Rev. **B58,** 6026 (1998)
[18] S. Ramanathan and D. Fisher, Phys. Rev. Lett. **79,** 877 (1977).